\documentclass{iau}
\usepackage{graphicx}

\title[Nature of the solar dynamo at small scales] 
{Nature of the solar dynamo at small scales}

\author[J.O. Stenflo]   
{J.O. Stenflo$^{1,2}$}

\affiliation{$^1$Institute of Astronomy, ETH Zurich, CH-8093 Zurich, Switzerland\\[\affilskip]
$^2$Istituto Ricerche Solari Locarno, Via Patocchi, CH-6605 Locarno Monti, Switzerland \\email: {\tt stenflo@astro.phys.ethz.ch}}

\pubyear{2013}
\volume{294}  
\jname{Solar and Astrophysical Dynamos and Magnetic Activity}
\editors{A.G. Kosovichev, E.M. de Gouveia Dal Pino, \& Y.Yan, eds.}
\begin{document}

\maketitle

\begin{abstract}
It is often claimed that there is not only one, but two different
types of solar dynamos: the one that is responsible for the
appearance of sunspots and the 11-yr cycle, frequently referred to as
the ``global dynamo'', and a statistically time-invariant dynamo,
generally referred to as the ``local dynamo'', which is supposed to be 
responsible for the ubiquitous magnetic structuring observed at small
scales. Here we examine the relative contributions of these two qualitatively
different dynamos to the small-scale magnetic flux, with the
following conclusion: The local dynamo does not play a significant
role at any of the spatially resolved
scales, nearly all the small-scale flux, including the flux revealed by
Hinode, is supplied by the global dynamo. This conclusion is reached
by careful determination of the Sun's noise-corrected basal magnetic
flux density while making use of a flux cancellation function determined
from Hinode data. The only allowed range where there may be
substantial or even dominating contributions from a local dynamo seems
to be the scales below 
about 10\,km, as suggested by observations of the Hanle depolarization effect in atomic
spectral lines. To determine the fraction of the Hanle
depolarization that may be due to the action of a local dynamo, a synoptic program is 
being initiated at IRSOL (Istituto Ricerche Solari Locarno). 
\end{abstract}

\firstsection 

\section{Introduction}\label{sec:intro}
The Sun's dynamo that is responsible for all solar activity with its
11-yr cycle depends on all scales for its operation. It is the
statistical left-right symmetry breaking of the convection at all
scales by the Coriolis force
induced by the Sun's rotation that causes the global, migrating
magnetic patterns that characterize the 11-yr cycle. Convective
turbulence, a fundamental ingredient of the 
dynamo, breaks up the generated large-scale structures such that they
cascade down the scale spectrum to the magnetic diffusion limit,
at scales of 10-100\,m, where the field lines cease to be frozen-in
and decouple from the plasma. As the symmetry breaking that generates
the global pattern takes place and contributes throughout the entire
scale range, it is not meaningful to conceptually divide this dynamo
into a large-scale and a small-scale part. All scales are
interconnected. 

From theory and numerical simulations it has however been suggested
that, besides the dynamo that is responsible for all of solar
activity, there is a second, qualitatively different dynamo that also
operates on the Sun and is responsible for the statistically
time-invariant small scale structuring of the magnetic field
(\cite{stenflo-petrovay93}; \cite{stenflo-cattaneo99}; \cite{stenflo-vs07}).  

The dynamo that is responsible for solar and stellar activity is
generally referred to as the ``global dynamo'', to distinguish it from
the time invariant, quiet-sun dynamo, which is generally referred to
as the ``local'' or ``surface'' dynamo. In the present paper we will
determine upper, empirical limits to the 
flux contributions from a local dynamo and thereby show that it is the
global dynamo, not the local one, which dominates the observed spatial
structuring, even at the smallest of all the resolved scales. 

These upper limits represent the ``basal'' magnetic flux density, the
minimum average unsigned flux density that can be found in the most
quiet regions on the Sun. As however these low average flux densities
are completely swamped by measurement noise in the MDI and HMI
magnetograms, we have to develop an accurate procedure to determine 
the noise and remove its influence on the
average unsigned flux densities. Fortunately the circumstance that the
noise and the intrinsic solar flux densities obey entirely different
scaling laws enables us to solve this problem in a well-defined
way. For high precision this also requires the use of the empirical, noise-corrected PDF
(probability density function) for the flux densities. Both the
magnetic field scaling law and the PDF have previously been determined from
analysis of Hinode quiet-sun data (\cite{stenflo-s10aa}, 2011). The
empirical scaling law is valid 
over a range that includes the Hinode, HMI, and MDI scales, thus allowing
the average unsigned flux densities to be translated across this
entire range without the need for any extrapolation.

\section{The Sun's magnetic scale spectrum}\label{sec:enspec}

\begin{figure}[t]
\begin{center}
\includegraphics[scale=0.35,angle=90.]{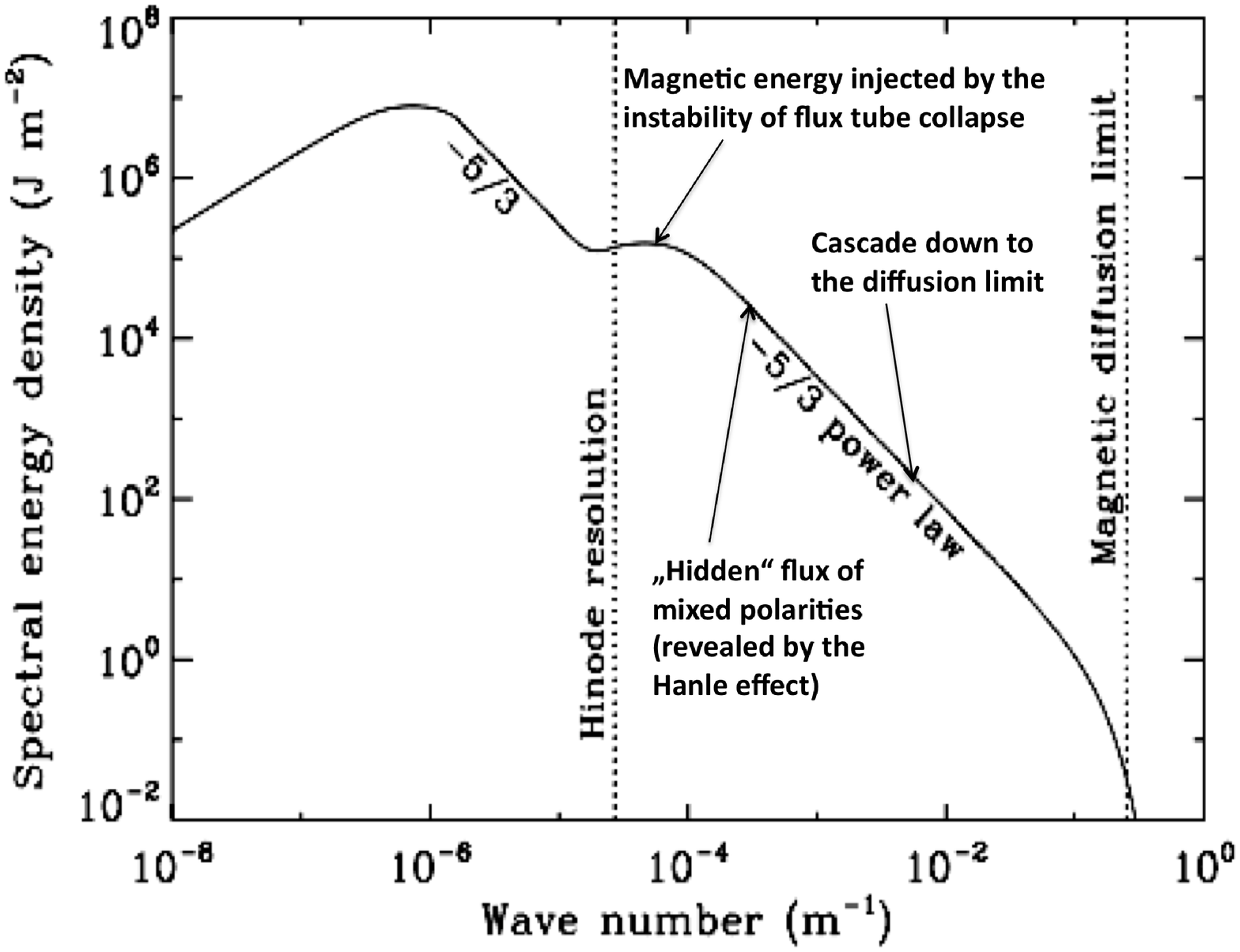}
\vspace*{-0.4 cm}
\caption{Magnetic energy spectrum of the quiet Sun over a scale range
  of more than 7 orders of magnitude (adapted from \cite{stenflo-s12aa1}).}\label{fig:enspec}
\end{center}
\end{figure} 

Figure \ref{fig:enspec} gives an overview of the magnetic energy
spectrum for the quiet Sun, spanning more than 7 orders of
magnitude. For the spatially resolved scales the 
spectrum has been determined from direct analysis of Hinode quiet-sun
data, while its continuation into the unresolved domain has been
inferred by indirect empirical techniques. In particular the bump in
the spectrum just beyond the Hinode resolution limit has been inferred
from analysis of Hinode line-ratio data that are interpreted in terms
of a collapsed, kG-type flux tube population in the 10-100\,km range,
and by modeling the influence of this population on the magnetic
energy spectrum. The vast amount of ``hidden'' flux that has been
inferred from analysis of the observed Hanle depolarization effect in
atomic lines is expected to have its main contribution from the scales
just below the range where the flux tubes are believed to reside. 

It is a consequence of the turbulent nature of the solar
plasma that the magnetic energy spectrum spans all these scales in a
continuous manner. Large-scale structures are spontaneously broken
up into smaller structures in a cascade down to the end of the scale
spectrum. The all-pervasive Coriolis force breaks the left-right
symmetry to generate a non-zero average helicity at all scales, which
is the key process of the cyclic dynamo. 

The circumstance that there is
substantial magnetic structuring at very small scales is by
itself no evidence for the existence of a local dynamo. Only the
part of this structuring that is time-invariant and decoupled from the
11-yr cycle could in principle have its origin in a local dynamo.

\section{Scaling laws and noise removal}\label{sec:scaling}
Analysis of Hinode magnetic-field data for the quiet Sun has shown
that the average unsigned vertical flux density $B_{\rm ave}$ obeys a scaling law
described by the {\it cancellation function} 
\begin{equation}
B_{\rm ave} \,\sim\, d^{-\kappa}\label{eq:canc}
\end{equation}
(\cite{stenflo-pietarila09}; \cite{stenflo-s11aa}), where $d$ is the
scale size (side of the square-shaped smoothing 
window), and $\kappa$ is the {\it cancellation exponent}. By numerical
smoothing of the Hinode data to simulate the effect of different
spatial resolutions, a cancellation function with a small,
approximately constant value of $\kappa$ across the whole range
from the Hinode 0.3\,arcsec scale up and beyond the MDI 4\,arcsec scale is
found. According to the analysis of \cite[Pietarila Graham et
al. (2009)]{stenflo-pietarila09} the numerical value of 
$\kappa$ is 0.26, while according to \cite[Stenflo
(2011)]{stenflo-s11aa} it is 
0.13. Both these values are much smaller than $\kappa$ for random 
noise, which is unity as a consequence of Poisson statistics. Due to
the great difference in $\kappa$ it becomes possible to achieve a
clean separation between the noise and the intrinsic solar contribution. 

\begin{figure}[t]
\begin{center}
\includegraphics[scale=0.35,angle=90]{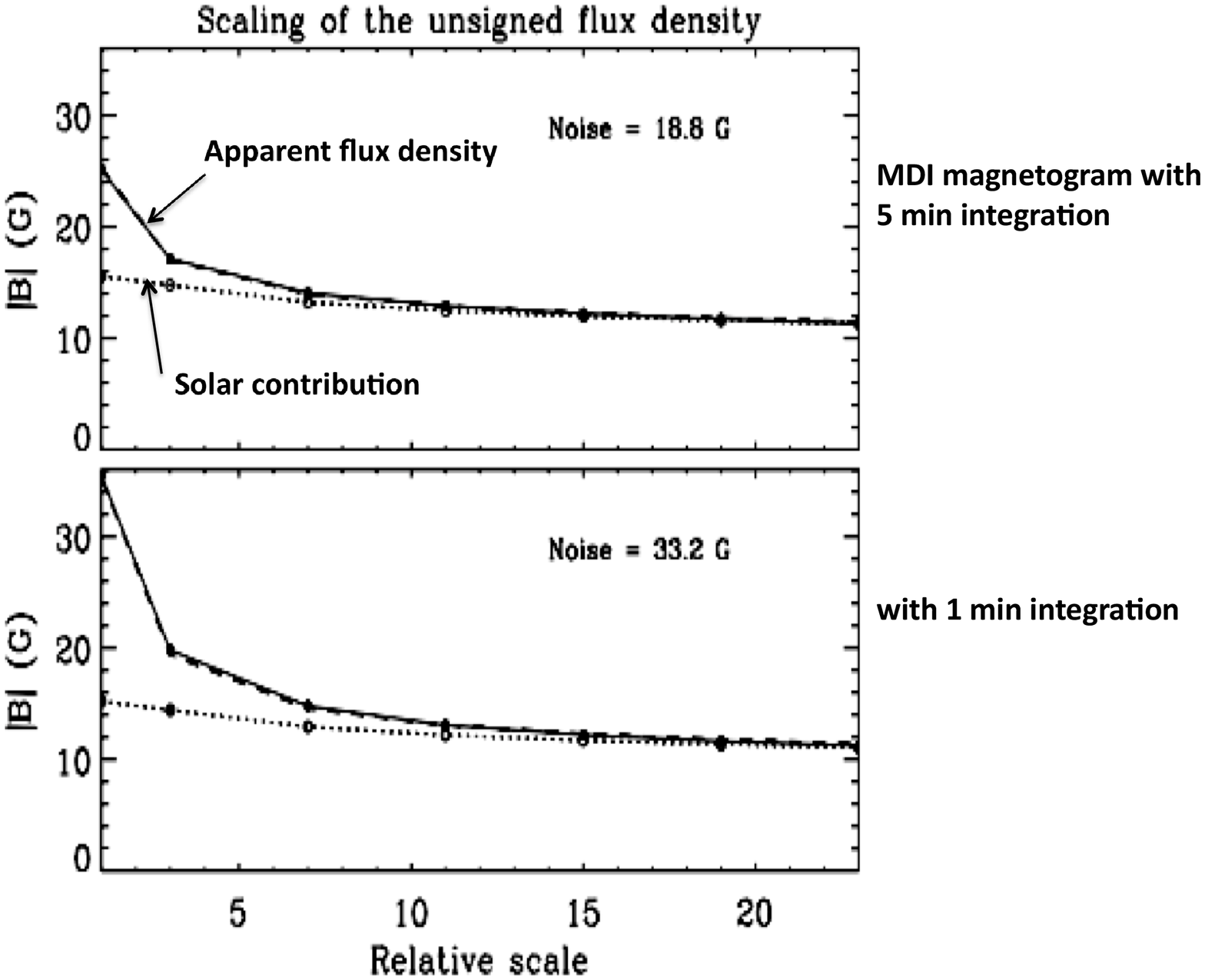}
\vspace*{-0.4 cm}
\caption{Illustration of the procedure for noise removal from the
  observed (apparent) average vertical flux densities, for the example
of two MDI magnetograms obtained with different integration
times. The relative scale of the horizontal axis is given in pixel
units. With a model that 
uses the greatly different scaling laws for  
the noise and the solar component, the noise-free average flux
density (dotted lines) can be determined from its noise-affected
counterpart (solid curves). With the fitted values for the noise $\sigma$ of 18.8 and
33.2\,G, respectively, the model (dashed curves) reproduces the
observed curves nearly perfectly.}\label{fig:mdiscale}
\end{center}
\end{figure}

As the size of the smoothing window is increased, the observed,
noise-affected apparent 
flux density $B_{\rm app}$ initially decreases steeply due to the
steep scaling law of the noise component, but subsequently levels out to
asymptotically approach the much shallower slope of the
solar scaling law. This behavior is illustrated in
Fig.~\ref{fig:mdiscale} for $B_{\rm app}$ (solid lines) as determined
from the disk-averaged vertical flux density in two different MDI
magnetograms, one recorded with 5\,min integration, the other with
1\,min integration. The dotted curves represent the intrinsic solar
components governed by a scaling law with cancellation
exponent 0.13. 

To accurately combine the solar and noise contributions we need to know
the empirical shape of the intrinsic (noise-free) PDF (probability
density function) for the vertical flux densities. This PDF has
previously been determined from analysis of Hinode quiet-sun data and
found to be characterized by an extremely narrow core peak,
represented by a stretched exponential centered at zero flux density,
and quadratically declining ``damping'' wings. 

The effect of noise is to convolve the solar PDF with the Gaussian
PDF of random noise. Through numerical Gaussian convolution of the
solar PDF, we have found that the relation between the apparent,
noise-affected average flux density $B_{\rm app}$ and its noise-free 
counterpart $B_{\rm ave}$ can be analytically represented by 
\begin{eqnarray}
B_{\rm ave} & = & [\,B_{\rm app}^\alpha\,-\,(0.798\,\sigma)^\alpha\,]^{(1/\alpha)}\,,\nonumber\\
\alpha & = & 1.36-0.004\,\sigma+0.0034\, B_{\rm app}\label{eq:noisemodel}
\end{eqnarray}
over the relevant parameter range.  $\sigma$ is the standard deviation of the Gaussian noise
distribution. The reason for the factor 0.798 is that the average
unsigned $x$ value of a Gaussian distribution with standard deviation
$\sigma$  is $0.798\sigma$. A practically identical relation is obtained if
we replace our solar PDF with a 
Lorentzian function, which shows that the detailed shape of the solar PDF
core region is unimportant in this context. 

Applying this noise model to all the 73,838 SOHO/MDI 96\,min cadence
magnetograms, we obtain excellent model fits to the observed average
unsigned flux density, as illustrated in Fig.~\ref{fig:mdiscale} by
the nearly perfect agreement between the solid (observations) and
dashed (model) curves. The model has two free parameters to
be determined by the fit: the noise $\sigma$, and the value of the
noise-free $B_{\rm ave}$ in the unsmoothed magnetogram. With a fixed
value of 0.13 for the cancellation exponent $\kappa$ that governs the
$B_{\rm ave}$ scaling (while for random noise, $\kappa$ is always
$=1$), the same values of $\sigma$ are found from 
the model fit for all the magnetograms that
have the same integration time (which is either 1 or 5\,min), regardless of
the value of $B_{\rm ave}$, which varies greatly
with the phase of the solar cycle. With other choices for
$\kappa$ the value of $\sigma$ becomes dependent on $B_{\rm ave}$,
which is unphysical, since $\sigma$ should represent measurement noise
that is independent of the conditions on the Sun. It is gratifying
that the value of $\kappa$ that is so demanded by physical consistency
is the same as previously determined by an entirely independent type
of analysis of Hinode quiet-sun data (\cite{stenflo-s11aa}). This
consistency is further confirmation of the validity of this scaling
law over the range that spans the Hinode and MDI scales. 

A more detailed account of the noise removal procedure and its
application to the MDI and HMI data for the determination of the basal
flux density can be found in \cite[Stenflo (2012b)]{stenflo-s12aa2}.

\section{Basal flux density from MDI analysis}\label{sec:mdi}
For each of the MDI magnetograms that span the 15-yr period May 1996
-- April 2011, we have converted the directly measured line-of-sight
component of the magnetic flux density to vertical flux density
through division by $\mu$, the cosine of the heliocentric angle,
assuming that on average the field is oriented in the vertical
direction due to the strong buoyancy forces acting on the kG type flux
tubes, which carry most of the flux in the photosphere. We then average these vertical
flux densities over all the pixels, but exclude a limb zone with a
width of 10\,\%\ of the disk radius. This average defines our
noise-affected value of $B_{\rm app}$. It is then converted to its
noise-free counterpart $B_{\rm ave}$ with our noise
removal model. 

\begin{figure}[t]
\begin{center}
\includegraphics[scale=0.35,angle=90]{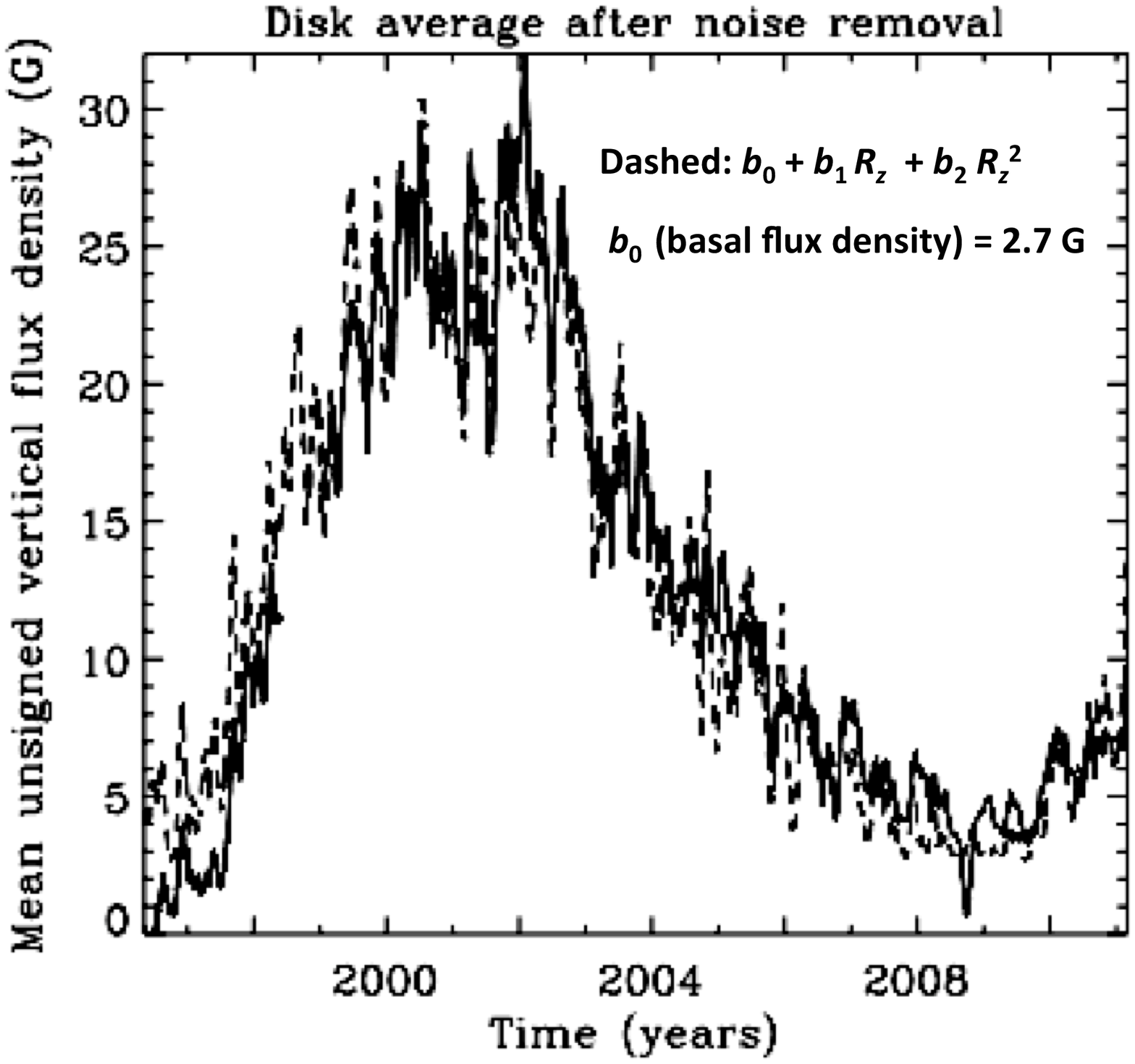}
\vspace*{-0.4 cm}
\caption{Noise-free, disk averaged unsigned vertical flux density
  $B_{\rm ave}$ (solid curve), compared with a second-order polynomial
of the sunspot number (dashed curve). The nearly perfect correlation
allows us to interpret the constant polynomial coefficient $b_0$ as
the basal vertical flux density, representing its value in the absence
of solar activity and sunspots.}\label{fig:sunspot}
\end{center}
\end{figure} 

Figure \ref{fig:sunspot} shows as the solid line the so determined
values of $B_{\rm ave}$ as a function of time for the 5-min
integration MDI magnetograms for the whole period
covered by the data set (excluding a data gap during 1998.4 -
1999.0). The corresponding results for the 1-min MDI magnetograms are
not illustrated here, because they look the same except for their
larger noise fluctuations. Since the solid curve has a striking resemblance to the
corresponding curve for the sunspot number $R_z$, we have tested the
sunspot correlation by fitting $B_{\rm ave}$ to a second-order
polynomial $b_0 +b_1 R_z +b_2 R_z^2$ in the sunspot number. This test
reveals a nearly one-to-one correspondence between $B_{\rm ave}$ and
the sunspot number, as shown by the dashed model curve in
Fig.~\ref{fig:sunspot}. 

Since $B_{\rm ave}$ can be modeled so closely by the sunspot number,
we may interpret the constant, cycle-independent term $b_0$ of the
polynomial fit as the value of $B_{\rm ave}$ in the absence of
sunspots, in other words, as the {\it basal} unsigned flux
density. This basal value is found to be 2.7\,G for the set of 5-min
integration MDI magnetograms.

\section{Basal flux density from HMI analysis}\label{sec:hmi}
The noise-removal model that we used for the MDI
magnetograms applies equally well to the SDO/HMI data. As the HMI  data set
so far mainly covers the deep minimum phase of the solar cycle and a
correlation with sunspots therefore cannot be done, it is not meaningful to do
averaging over the whole solar disk. Instead we have for HMI defined
the noise-affected average unsigned flux density $B_{\rm app}$ as the
average over the innermost 10\,\%\ of the radius $r_\odot$,
i.e., $r/r_\odot <0.1$. Since $\mu\approx 1.0$ over this whole region,
the line-of-sight component of the flux density equals the 
vertical component, no conversion is needed. 

When applying our noise model to the HMI disk center data with
cancellation exponent $\kappa=0.13$, we find a value of the noise
$\sigma$ (8.0\,G) that is independent of $B_{\rm ave}$, as required by
physical consistency. Other choices of $\kappa$ do not have this
property, which again supports the validity of this scaling law.  

\begin{figure}[t]
\begin{center}
\includegraphics[scale=0.32,angle=90]{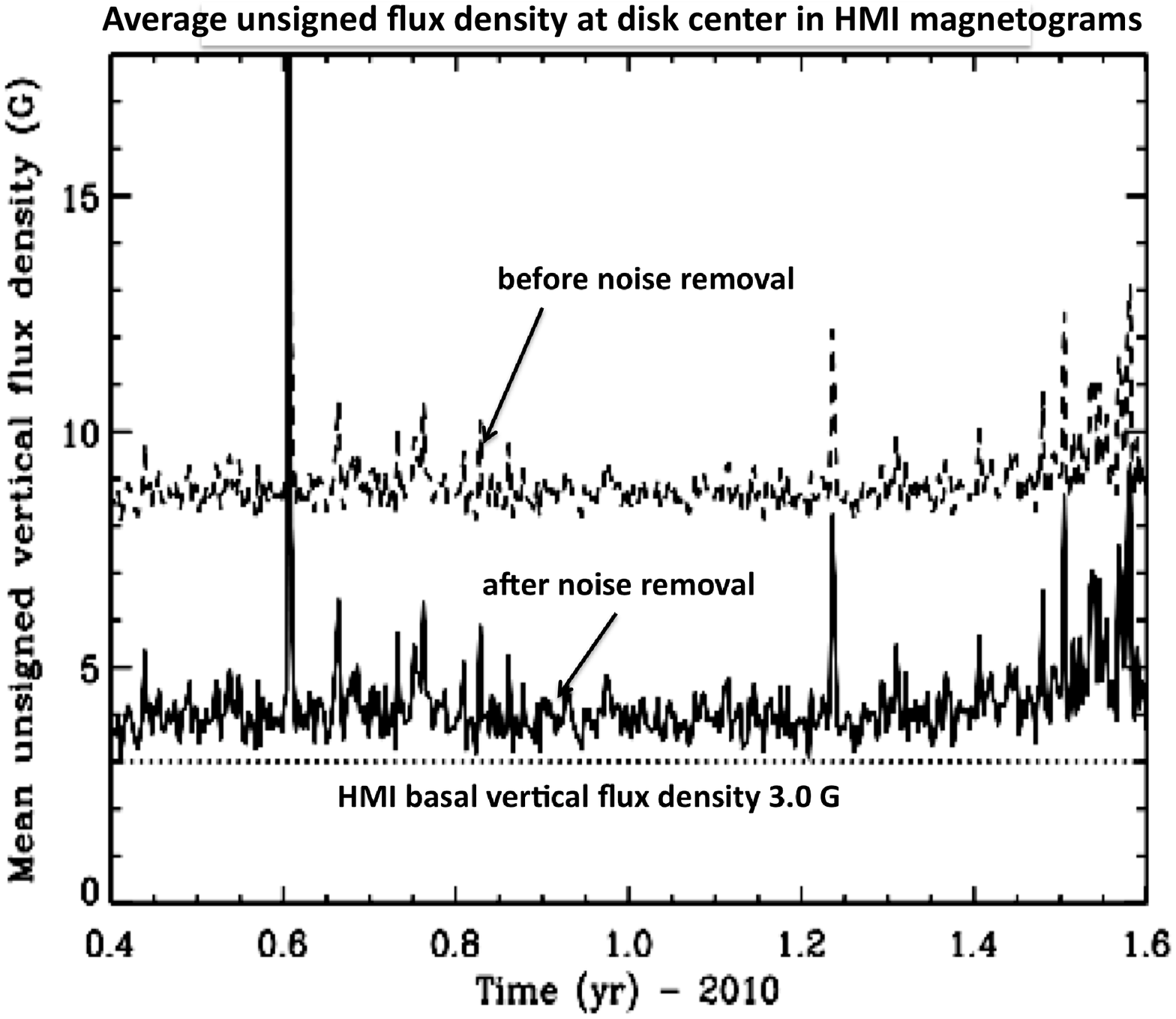}
\vspace*{-0.4 cm}
\caption{Time series of the noise-affected $B_{\rm app}$ (dashed
  line), defined as the average unsigned vertical flux density within
  the disk center region $r/r_\odot <0.1$ of the HMI magnetograms,
  here selected with a cadence of 24\,hr. When noise is removed $B_{\rm app}$ becomes
  $B_{\rm ave}$, shown as the solid line. Its lower envelope is
  3.0\,G, which represents the basal flux density for the HMI spatial
  scale.}\label{fig:hmiseries}
\end{center}
\end{figure} 

In Fig.~\ref{fig:hmiseries} the directly observed, noise-affected
$B_{\rm app}$ is given as a function of time for the period 2010.4 -
2011.6 as the dashed curve, while its noise-corrected counterpart $B_{\rm
  ave}$ is the solid curve. Their spiky appearance is not due to noise
but is solar, due to the circumstance 
that we have only selected one HMI magnetogram per
24\,hr. Since solar rotation brings significantly different magnetic
features into the analysed disk center region, $B_{\rm app}$ becomes
jumpy when using such course time
cadence. These solar spikes get mapped one by one when converting the dashed
curve into the solid one, the conversion procedure does not create
additional fluctuations. 

The basal vertical flux density can now be obtained simply as the
lower envelope to the solid curve, 3.0\,G , below which the average flux density 
never dips, even in the most quiet solar regions during this record
deep cycle minimum phase. 

Using our cancellation function with $\kappa=0.13$, we can translate
the HMI basal flux density to its corresponding value at the MDI
scale. Since the spatial resolutions of HMI and MDI differ by a factor
of 4, the 3.0\,G HMI value translates to $3.0\times
4^{-0.13}\,=2.5$\,G, which is in nearly perfect agreement with the
2.7\,G value that was based on the correlation of the 5-min integration
MDI magnetograms with the sunspot number. 

Similarly we can translate the HMI basal flux density from the 1\,arcsec
HMI scale to the Hinode
0.3\,arcsec scale, since our cancellation function was derived from
Hinode data and later validated by the HMI and MDI data. This gives $3.0\times
0.3^{-0.13}\,=3.5$\,G, a small number compared with the average
unsigned flux densities of typically 10-15\,G found from Hinode
analysis of quiet regions (\cite[Lites et al. 2008]{stenflo-litesetal08};
\cite{stenflo-s11aa}), indicating that the quiet regions that 
were analysed with Hinode were not representative of the basal flux level but
contained substantially more flux. This issue will be examined in the
next section.

\section{Variations of quiet-sun magnetic fields}\label{sec:qsun}
During the period covered by our HMI analysis there were no active
regions near disk center, the fluctuations that we see from day to
day in Fig.~\ref{fig:hmiseries} represent quiet regions with varying average
flux densities. Such variations in the amount of flux can only come
from the global dynamo, since the local dynamo has nothing to do with
the solar cycle and therefore is time invariant. 

The circumstance that there is a non-zero basal flux density in
the absence of any sunspot activity does not imply that the
local dynamo must be the source of this basal flux. Instead it is to
be expected that the large amounts of flux that have been generated by
the global dynamo during the course of the 11-yr cycle do not
instantly disappear when the solar activity becomes zero, but that
flux remnants will linger around for extended periods of time and only
slowly decay asymptotically towards zero if there is no supply of new
flux. For this reason the determined non-zero basal flux density only defines an
{\it upper limit} to the contributions from a hypothetical local
dynamo, but not evidence for its existence. 

\begin{figure}[t]
\vspace*{-0.2 cm}
\begin{center}
\includegraphics[scale=0.45,angle=90]{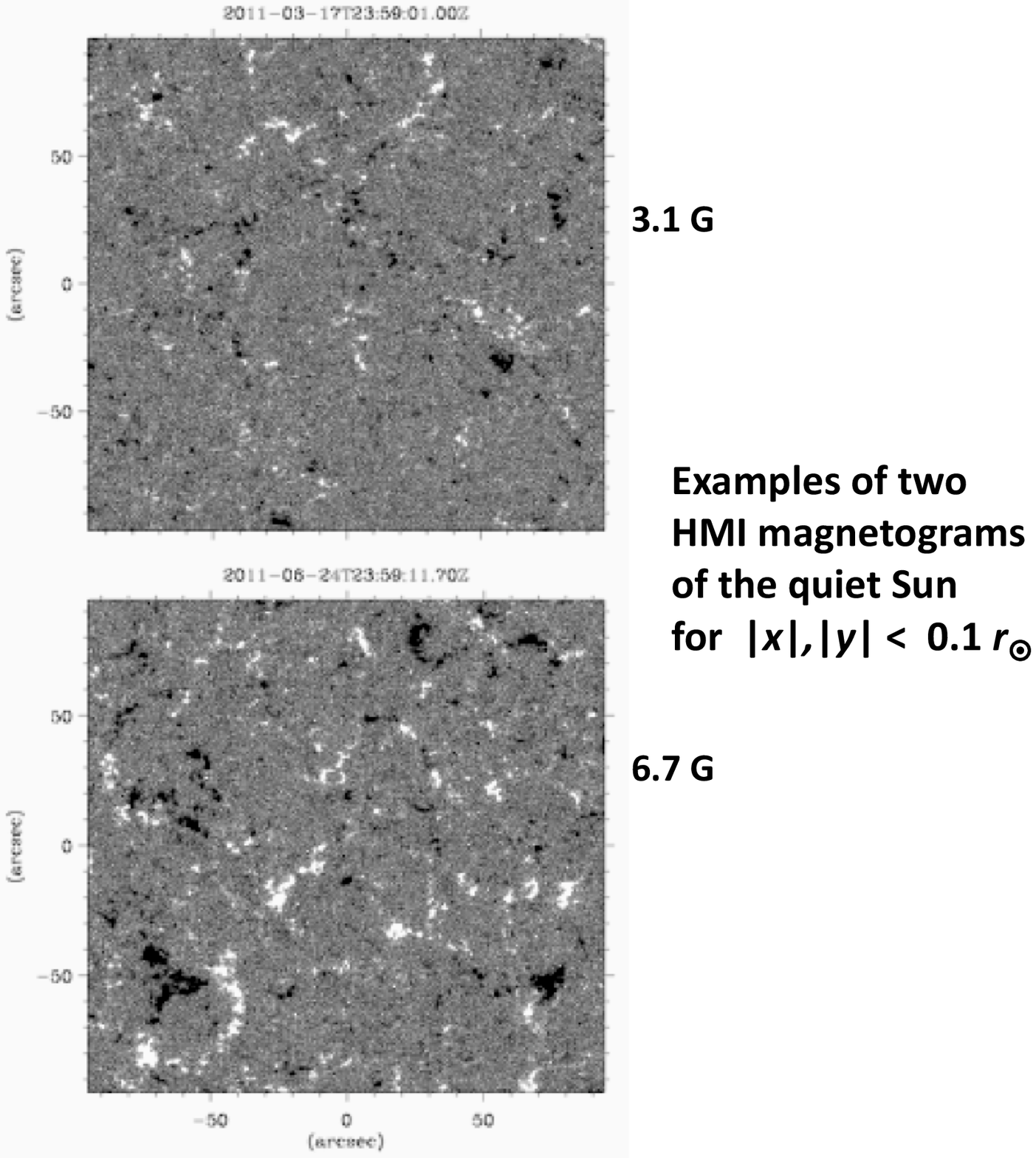}
\vspace*{-0.8 cm}
\caption{Comparison between the quiet-sun disk-center regions of two HMI
  magnetograms, obtained at times 2011.2086 (upper panel) and
  2011.4796 (lower panel), representing regions with
  basal flux density and with 2.2 times the basal flux
  density, respectively. The grey-scale cuts ($\pm 50$\,G) are the
  same for the two panels.}\label{fig:2dplots}
\end{center}
\end{figure} 

The validity of the conclusion that the noise-corrected HMI basal
flux density is approximately 3\,G, and not significantly less (like
zero) or larger, can easily be verified by direct visual
inspection of the corresponding magnetograms and by identifying the 
parts of the PDF that contribute to 
the bulk of the observed flux. For this purpose we have selected the
disk-center magnetograms from two days, one representing the basal,
minimum average flux density (upper panel of Fig.~\ref{fig:2dplots},
recorded at time 2011.2086), the other having an average flux density
more than twice as large (6.7\,G, lower panel of Fig.~\ref{fig:2dplots},
recorded at time 2011.4796). While it is obvious that there is much
more flux in the quiet region of the lower panel, it is striking that
the qualitative appearance of the patterns is very similar in the two
panels. Thus the basal flux region exhibits the same kind of strongly
intermittent pattern that is network-like on the supergranulation scale. 

Before noise correction, the average unsigned flux densities of the
two regions were 8.2 and 10.8\,G, thus in proportion 1 : 1.3, which is a
small difference in comparison with the proportion 3.1 : 6.7 = 1 : 2.2
after noise correction. If the basal flux density were much closer to
zero than 3\,G, then the above proportion would escalate, and we would
not see as much magnetic structures in the upper panel as we do
see. This visual check confirms that our non-zero basal flux density
cannot be very different from our determined value. 

\begin{figure}[t]
\begin{center}
\includegraphics[scale=0.33,angle=90]{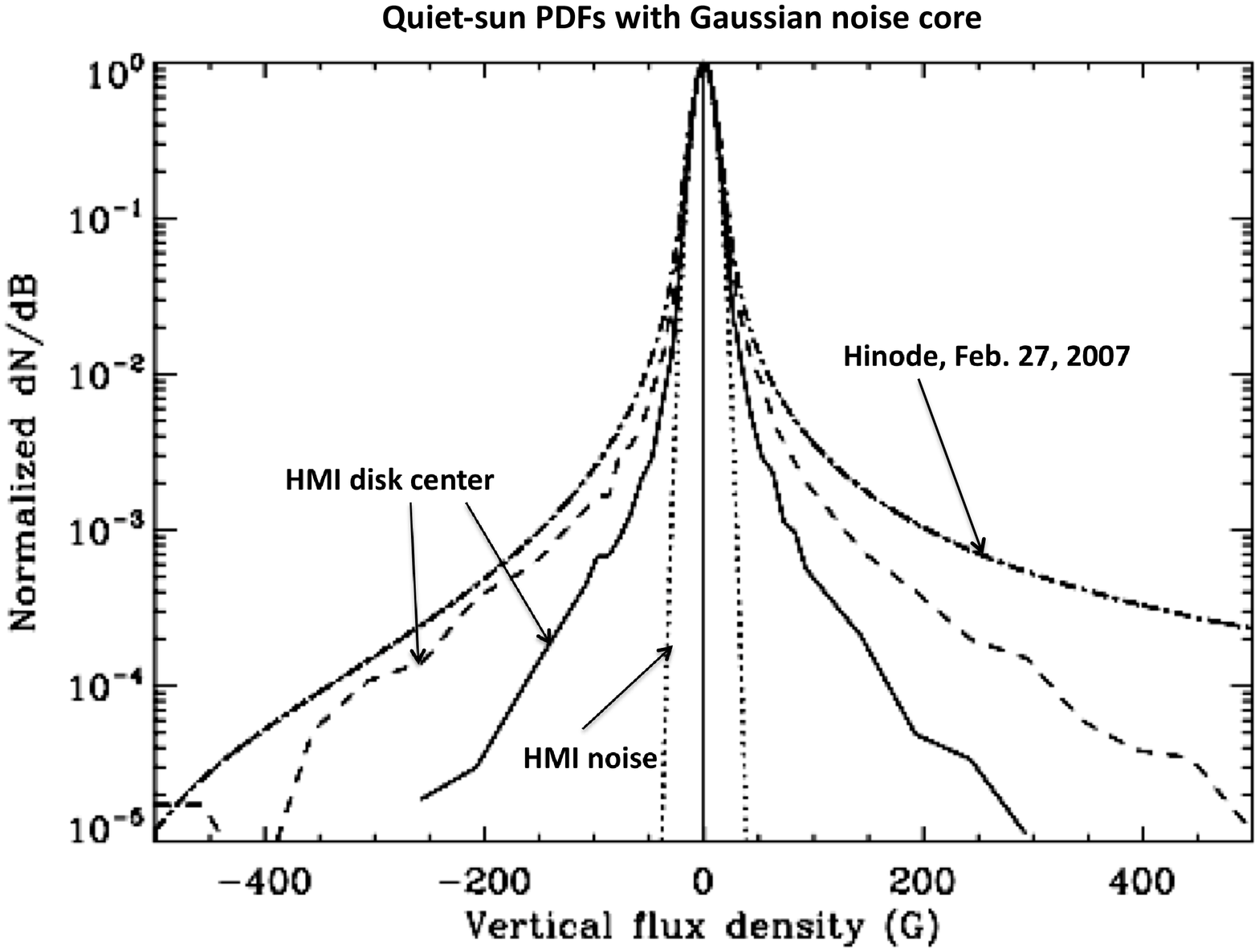}
\vspace*{-0.4 cm}
\caption{Amplitude-normalized probability density functions (PDF) for
  the upper and lower magnetogram of Fig.~\ref{fig:2dplots} (solid and
dashed curve, respectively), for the HMI noise distribution (dotted
curve), and for the PDF that has been determined from Hinode observations of
February 27, 2007 (\cite{stenflo-s10aa}), after noise removal and subsequent convolution
with the HMI noise Gaussian, to allow comparison of the different PDF
wings on the same scale. While the solid curve is representative of
the basal flux density, everything above it expresses the cycle
dependence of the background quiet-sun flux.} \label{fig:pdf}
\end{center}
\end{figure} 

To further clarify the role of noise and the entirely different nature
of the intrinsic solar magnetic pattern we show in Fig.~\ref{fig:pdf} the
PDFs of the two magnetograms of Fig.~\ref{fig:2dplots}, as the solid line
for the basal magnetogram, and as the dashed line for the magnetogram with
2.2 times the basal flux density. These PDFs are
uncorrected for noise, which means that they are smeared by
convolution with the $\sigma=8.0$\,G Gaussian noise 
distribution, which is plotted separately as the
dotted line. The circumstance that this Gaussian is practically
indistinguishable from the shapes of the core regions of the solid and
dashed PDFs implies that the width of the
intrinsic (noise-free) PDF core peak is much smaller than the 8\,G of
the noise Gaussian. 

The much slower, approximately quadratical decline of the PDF wings
implies that they are nearly unaffected by the noise convolution
and therefore represent the wing shape of the intrinsic
PDFs. Practically all the noise-corrected average unsigned flux
density comes from the PDF wings, the contribution from the core
region is insignificant due to the extremely narrow intrinsic width of the core
peak. Before noise correction, however, the average unsigned flux
density is dominated by the contribution from the core region, where
all the noise resides. 

The extended PDF wings are a signature of spatial {\it intermittency},
since the high flux densities in the wings combined with the low
occurrence probability implies that the corresponding flux elements
are relatively rare, and that they must therefore be statistically well
separated (i.e., intermittent). Since the bulk of the flux comes from
the PDF wings, this implies that most of the flux is highly
intermittent, which confirms the visual impression that we get from the
magnetograms in Fig.~\ref{fig:2dplots}. As all the noise in the
plotted magnetograms has values far below the chosen grey-scale cuts
of $\pm 50$\,G, the noise is inconspicuous. The visual impression of the
pattern comes exclusively from the contributions of the PDF
wings. Still it is the inconspicuous noise background that dominates
the apparent, average unsigned flux density $B_{\rm app}$. 

For comparison we show as the dash-dotted line in Fig.~\ref{fig:pdf}
the analytical representation of the noise-free PDF derived from Hinode SOT/SP
observations of the quiet-sun disk center on February 27, 2007
(\cite{stenflo-s10aa}). However, before plotting we have convolved it with
the $\sigma=8.0$\,G HMI Gaussian noise distribution, because only when
all the PDFs in the figure have been smeared with the same noise
function and been amplitude normalized, it becomes possible to compare the levels of
the different PDF wings with each other on a common scale. We see that
the Hinode PDF wings are substantially more elevated than the PDF
wings of our two HMI magnetograms and therefore contribute much more
flux. The main reason why analysis of Hinode quiet-sun data has
indicated much higher flux densities than what one can expect from our
small basal flux density is simply because the observations were not
carried out during the deepest part of the solar activity
minimum. Therefore there was much more background magnetic flux 
on the quiet Sun. Due to the shallowness of the
cancellation function the elevated Hinode PDF wings cannot be
explained in terms of scaling between the 1 and 0.3\,arcsec scales. 

For this reason we can safely say that the bulk of the quiet-sun
vertical magnetic flux observed with Hinode cannot have its origin in a local
dynamo, since the Hinode flux densities much exceed the basal
flux density.

\section{Basal flux density of the horizontal fields}\label{sec:horizontal}
The Hinode SOT/SP instrument has the capability of measuring all Stokes
parameters and therefore in principle of determining the magnetic
field vector. The polarimetric precision is high: for the circular
polarization (Stokes $V$) it is 0.047\,\%\ (in units of the continuum
intensity), for the linear polarization it is even better, 0.035\,\%\
in each of Stokes $Q$ and $U$ (\cite{stenflo-s11aa}). For those who are not so
familiar with the profound difference between the longitudinal and the
transverse Zeeman effects, the high polarimetric sensitivity for the
linear polarization may give the completely misleading
impression that transverse and longitudinal magnetic fields can be
determined with comparable accuracy. 

The approximate conversion laws between the Stokes $V$, $Q$, and $U$ amplitudes (in
units of the continuum intensity) and the longitudinal and transverse flux
densities $B_\parallel$ and $B_\perp$ for field strengths below about
0.5\,kG, assuming for simplicity that the field is resolved (filling
factor unity), are 
\begin{eqnarray}
B_\parallel \,{\rm (G)}\,& = &\,29.4\,V{\rm (\%)}\,,\nonumber\\
B_\perp \,{\rm (G)}\,& = &\,184\,[\,Q{\rm (\%)}^2 +U{\rm (\%)}^2\,\,]^{1/4}\,,\label{eq:weakfield}
\end{eqnarray}
where the proportionality factors have been determined by
radiative-transfer modeling of the Fe\,{\sc i} 6301\,\AA\ line used by
Hinode (\cite{stenflo-s11aa}). We note that while the relation between field
strength and polarization is linear in the longitudinal case, it is
highly non-linear with a large proportionality factor in the
transverse case. 

\begin{figure}[t]
\begin{center}
\includegraphics[scale=0.45]{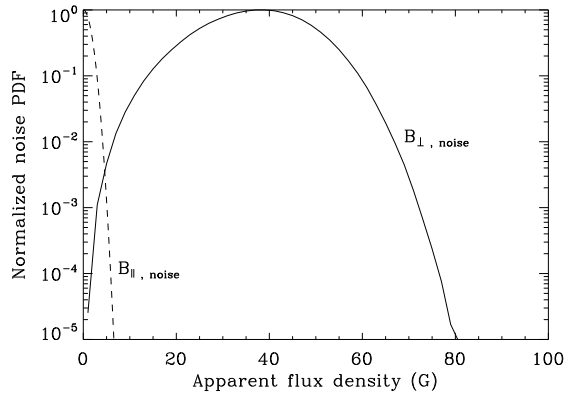}
\caption{PDFs of the longitudinal and transverse apparent flux densities, 
  exclusively due to Gaussian polarimetric noise in the Hinode SOT/SP
  recordings of the Fe\,{\sc i} 6301\,\AA\ line. As the noise PDF for
  the horizontal flux density is ubiquitous and time invariant, it
  masquerades as a basal flux density.}\label{fig:bnoise}
\end{center}
\end{figure} 

With these equations we can use Monte Carlo techniques to
convert the Gaussian noise distributions in Stokes $V$ (with
$\sigma=0.047$\,\%) and in $Q$ and $U$ (each with $\sigma=0.035$\,\%)
into PDFs for the apparent magnetic flux density. The result is shown
in Fig.~\ref{fig:bnoise} (same as was plotted on a linear scale in
Fig.~A.1 of \cite{stenflo-s11aa}). If these PDFs would mistakenly be assumed to
represent solar properties, they would suggest that the Sun is
completely swamped by horizontal magnetic flux with typical flux
densities of order 40\,G, with a wide distribution of field strengths. In comparison,
the contribution of the vertical fields would seem insignificant. 

The PDF for the directly determined apparent flux densities, which is
composed of a combination of the noise and solar contributions, will
always be extended to larger flux densities than the noise PDFs
alone. Thus the distributions in Fig.~\ref{fig:bnoise} represent 
ubiquitous, time-invariant, lower limits to the observed PDFs, since the
instrumental noise hardly changes and at least does not improve over
the Hinode mission. The noise PDF for $B_\perp$ therefore 
masquerades as a huge basal flux density, although it is nothing of the
sort. 

If one wants to make a claim from observations that horizontal fields
are of significance for the local dynamo, then one must first deal with the
formidable problem how to remove the colossal noise contribution 
from the apparent horizontal flux density. In the previous sections we have
addressed the noise removal problem for the line-of-sight flux
density, but because of the approximately 25 times larger noise level
of the transverse field (after conversion to G units), and most of all
because of its non-linear conversion behavior, the
noise problem for the horizontal fields is of a completely different
magnitude. 

There have been repeated claims that Hinode data have revealed the
existence of much more horizontal than vertical flux in the quiet-sun
photosphere, or in other words, that the angular distribution of the field vectors
is much more pancake like than isotropic. Related to this are also
claims that the vast amounts of such 
horizontal flux must be due to a local dynamo, since it is always
present. Unfortunately none of these claims is based on any serious
attempt to address the formidable noise removal problem for the
transverse fields. 

When the noise issue has been seriously addressed in determinations of
the angular distribution of the field vectors, it is found that
instead of being pancake like, the distribution favors the vertical
direction, but becomes increasingly isotropic as we approach the
limit of small flux densities (\cite{stenflo-s10aa}). This implies that there
is more vertical than horizontal flux on the quiet Sun, and that there
is no valid evidence for the existence of a local dynamo from
observations of horizontal magnetic fields.

\section{A role for the local dynamo at scales below 10\,km}\label{sec:hanle}
Our low basal vertical flux density of 3.0\,G determined from analysis
of HMI data, which corresponds to 3.5\,G at the Hinode resolution,
implies that the bulk of the magnetic structuring seen by Hinode in
quiet solar regions has nothing to do with a local dynamo but
represents background magnetic flux that has been generated by the
global dynamo. The mentioned basal flux densities represent upper
limits to the contributions of a hypothetical local dynamo, but not
evidence for its existence, since we expect that there will always
remain some long-lived left-over non-zero background flux when the
solar activity with its supply of new flux from the Sun's interior
temporarily disappears. Our low upper limit therefore indicates that
the local dynamo does not play any significant role on the Sun at any
of the spatially resolved scales, including the Hinode scale, and that
the small-scale magnetic flux that we see has instead been produced by the
global dynamo. 

This does not rule out the possibility that the local dynamo can
become significant at smaller, still unresolved
scales. The only known way to empirically address this issue is
through observations of the Hanle effect, through its depolarizing effect on the
spectral-line polarization caused by coherent scattering. The observed
depolarization can be converted into a turbulent field strength (\cite{stenflo-s82}), which
is unobservable with the Zeeman effect due to cancellation of the 
contributions from 
unresolved opposite polarities. Such cancellations do not occur for
the Hanle effect due to its different symmetry properties. 

While the Hanle effect in optically thin molecular lines gives us a
turbulent field that is too weak (of order 10\,G) to point towards the
significant role of a local dynamo, the Hanle effect in atomic lines, 
in particular the Sr\,{\sc i} 4607\,\AA\ line
(\cite{stenflo-trujetal04}) and in the optically thick
molecular CN lines (\cite{stenflo-shapiro11}), indicates much stronger
turbulent fields, of order 
60\,G or more, which is too large to be reached by a scaling law
that connects with the resolved flux densities supplied by the global
dynamo. This strongly suggests a dominating role for the local
dynamo at the scales responsible for the Hanle depolarization signatures. 

The apparently contradictory low Hanle field strengths found with the
optically thin molecular lines can be given a
consistent explanation if most of the strong fields are located in the
intergranular lanes, which do not contribute much to the formation of
these molecular lines. This explanation however still needs to
be verified by future observations of the Hanle effect with high
spatial resolution, to allow the relative contributions from the
intergranular lanes and the cell interiors to be clearly separated from each other. 

Whether or not these strong turbulent fields have anything to do with
the local dynamo depends on how much of them are statistically time invariant,
independent of the phase of the solar cycle. This remains a
controversial issue. On the one hand the general appearance of the
Second Solar Spectrum (the linearly polarized spectrum that is
exclusively formed by coherent scattering processes) is found to vary
significantly with the solar cycle (\cite{stenflo-s1spw3}), and such variations
can only be magnetically induced (via the Hanle effect), indicating
substantial cycle variations of the ``hidden'' field. On the other
hand \cite[Trujillo Bueno et al. (2004)]{stenflo-trujetal04} have
compiled evidence from various observations of the 
Sr\,{\sc i} 4607\,\AA\ line that there is no significant variation of
the 60\,G type turbulent field with the cycle. 

The only known way to quantitatively resolve this issue is through a synoptic
program for the Hanle depolarization in atomic lines like the Sr\,{\sc
  i} line. A reason why this has not been done before is that
trustworthy synoptic results can only be obtained with a {\it
  differential} method, allowing us to normalize the recorded
scattering polarization amplitude to something that is independent of
the solar cycle. The problem has been that good Hanle lines like the
Sr\,{\sc i} 4607\,\AA\ line are spectrally isolated with no suitable 
reference lines close by, and that we cannot maintain the
polarization scale or the chosen limb distance of the spectrograph slit
to be sufficiently constant over time scales of the solar cycle. We have
however recently found an elegant solution to this problem, as follows: 

The line with the largest scattering polarization amplitude in the
whole visible spectrum is the Ca\,{\sc i} 4227\,\AA\ line. It is only
affected by the Hanle effect in its Doppler line core, while its
unblended prominent polarization peak in the blue line wing is immune
to the Hanle effect and therefore invariant with the solar
cycle. While this blue wing peak therefore can serve as an ideal
normalization reference for a synoptic program, its spectral location
is far from suitable photospheric Hanle lines like the Sr\,{\sc i} 4607\,\AA\
line. However, when the Sr line is observed in the 11th grating order
and the Ca line in the 12th order, they fall adjacent to each other 
in the spectral focus. With two square-shaped order-separating
interference filters side by side immediately before the CCD detector, the two
lines can be recorded simultaneously on the same CCD. 
This assures that both lines get recorded at the identical limb
distance and seeing conditions. The ratio between the Sr
polarization amplitude and the Ca blue wing amplitude is then a 
differential measure of the Hanle depolarization in the Sr
line on a scale that is stable over solar cycle time scales. A synoptic program
based on this technique will now be started at IRSOL (Istituto
Ricerche Solari Locarno) with the ZIMPOL-3 imaging Stokes
polarimeter. 

The spatial scales responsible for the observed Hanle 
depolarization signatures must be much smaller than the resolved
scales, otherwise much more of these ``hidden'' fields would already
be visible in magnetograms. Based on indirect arguments
(\cite{stenflo-s12aa1}) it is conjectured 
that the main contribution to these Hanle signatures comes from scales
that are below the scale range of 10-100\,km, where most of the
collapsed, kG-type flux tubes are expected to reside
(cf. Fig.~\ref{fig:enspec}). While the local 
dynamo appears to be insignificant at all the resolved scales, it
could dominate at scales below 10\,km. This would raise the
small-scale end of the magnetic energy spectrum and make it shallower
than the reference spectrum of Fig.~\ref{fig:enspec}. Current data from the Hanle
effect suggest that this might indeed be the case. With the synoptic
program that is being started at IRSOL we hope to 
determine the fraction of the observed Hanle depolarization signatures
that could be attributed to the action of a local dynamo.


\end{document}